\documentclass[
 aip,       
 amsmath,amssymb,
 twocolumn,  
]{revtex4}

\usepackage{textcase}
\usepackage{graphicx}
\usepackage{dcolumn}
\usepackage{bm}
\usepackage{amsmath,amsthm,amssymb,mathrsfs}

\usepackage{amsmath,amsthm,amssymb,mathrsfs}
\usepackage{graphicx} \usepackage{bm} \usepackage{amsmath}
\usepackage{amssymb}

\begin{document}

\title{Critical Field of Spin Torque Oscillator with Perpendicularly Magnetized Free Layer}

\author{Tomohiro Taniguchi} 
\author{Hiroko Arai} 
\author{Sumito Tsunegi} 
\author{Shingo Tamaru} 
\author{Hitoshi Kubota} 
\author{Hiroshi Imamura} 

\affiliation{ 
  National Institute of Advanced Industrial Science and Technology (AIST), Spintronics Research Center, 1-1-1 Umezono, Tsukuba 305-8568, Japan\\
}


\begin{abstract}
  The oscillation properties of a spin torque oscillator 
  consisting of a perpendicularly magnetized free layer 
  and an in-plane magnetized pinned layer 
  are studied based on an analysis of the energy balance between spin torque and damping. 
  The critical value of an external magnetic field applied normal to the film plane is found, 
  below which the controllable range of the oscillation frequency by the current is suppressed. 
  The value of the critical field depends on 
  the magnetic anisotropy, the saturation magnetization, and the spin torque parameter. 
\end{abstract}


\maketitle

The self-oscillation of the magnetization 
in a spin torque oscillator (STO) has been studied extensively 
because of its potential application to spintronics devices 
such as microwave generators and recording heads of high-density hard disk drives 
\cite{kiselev03,rippard04,houssameddine07,deac08,rippard10,kudo10,suto11,sinha11,zeng12,zeng13}. 
The self-oscillation of the magnetization is induced 
when the energy supply from the spin torque balances with 
the energy dissipation due to the damping \cite{slavin09}. 
Recently, it was found that 
the STO consisting of a magnetic tunnel junction (MTJ) 
with a perpendicularly magnetized free layer 
and an in-plane magnetized pinned layer \cite{yakata09,ikeda10,kubota12} 
showed a large power ($\sim 0.5$ $\mu$W) and 
a narrow linewidth ($\sim 50$ MHz) \cite{kubota13}, 
making a great advance toward the realization of STO devices. 


Precise control of the oscillation frequency by the current 
is necessary for STO application. 
To this end, 
it is important to clarify the oscillation properties of STOs. 
Depending on the magnetization directions of the free and pinned layers, 
STO can be classified into four types. 
The self-oscillation of the STO was first observed 
in the in-plane magnetized system in 2003 \cite{kiselev03}. 
An MTJ with 
an in-plane magnetized free layer and a perpendicularly magnetized pinned layer 
was also developed 
because a significant reduction in the switching current was expected \cite{houssameddine07,kent04,lee05}. 
An STO in which 
both the free and pinned layers were perpendicularly magnetized 
was theoretically studied 
because its axial symmetry made the analyses easy \cite{silva10}. 
Contrary to these three types, 
the oscillation properties of STO 
with a perpendicularly magnetized free layer 
and an in-plane magnetized pinned layer remain unclear. 


In this letter, 
we theoretically study the oscillation properties of an STO 
with a perpendicularly magnetized free layer 
and an in-plane magnetized free layer 
based on the analyses of the energy balance 
between spin torque and damping. 
We find that an external magnetic field 
applied normal to the film plane plays a key role 
in the self-oscillation of the magnetization. 
A critical value of the applied field exists 
below which the controllable range of 
the oscillation frequency by the current is suppressed. 
The value of the critical field depends on 
the perpendicular magnetic anisotropy, 
the saturation magnetization, 
and the spin torque parameter. 



\begin{figure}
\centerline{\includegraphics[width=1.0\columnwidth]{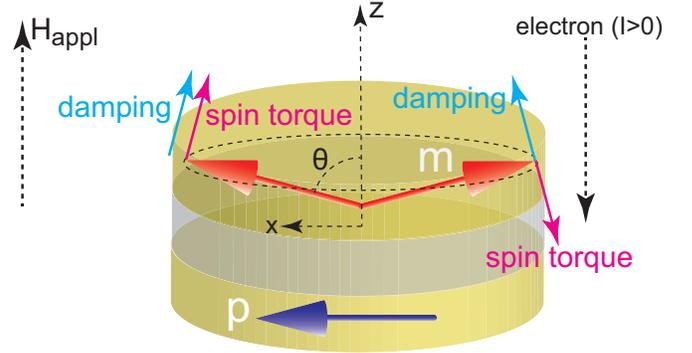}}\vspace{-3.0ex}
\caption{
         Schematic view of the system, 
         where $\mathbf{m}$ and $\mathbf{p}$ are the unit vectors pointing in the magnetization directions 
         of the free and pinned layers, respectively. 
         The tilted angle of the magnetization $\mathbf{m}$ from the $z$-axis is denoted as $\theta$. 
         The arrows indicate the directions of spin torque and damping. 
         The applied field is denoted as $H_{\rm appl}$. 
         \vspace{-3ex}}
\label{fig:fig1}
\end{figure}



The system we consider is schematically shown in Fig. \ref{fig:fig1}, 
where the unit vectors pointing in the magnetization directions of 
the free and pinned layers are denoted as 
$\mathbf{m}$ and $\mathbf{p}$, respectively. 
The $z$-axis is normal to the film plane 
while the $x$-axis is parallel to $\mathbf{p}$. 
The applied field, $H_{\rm appl}$, is parallel to the $z$-axis. 
The current is denoted as $I$, 
where the positive current corresponds to 
the electrons flowing from the free layer to the pinned layer. 
The magnetic energy density of the free layer, 
$E=-MH_{\rm appl}m_{z}-[M(H_{\rm K}-4\pi M)/2]m_{z}^{2}$, 
consists of the Zeeman energy 
and the uniaxial anisotropy energy, 
where $M$ and $H_{\rm K}$ are 
the magnetization and the crystalline anisotropy along the $z$-axis, respectively. 
Because we are interested in the perpendicularly magnetized free layer, 
$H_{\rm K}$ should be larger than the demagnetization field $4\pi M$. 
The energy density has two minima at $\mathbf{m}=\pm\mathbf{e}_{z}$. 
Throughout this letter, 
the initial state is assumed to be $\mathbf{m}=\mathbf{e}_{z}$. 
It should be noted that 
a trajectory with a constant $m_{z}=\cos\theta$ corresponds to 
the constant energy line of this system. 
The magnetization dynamics is 
described by the Landau-Lifshitz-Gilbert (LLG) equation \cite{lifshitz80,gilbert04,slonczewski89,slonczewski96,slonczewski02}, 
\begin{equation}
  \frac{d \mathbf{m}}{dt}
  =
  -\gamma
  \mathbf{m}
  \times
  \mathbf{H}
  -
  \gamma 
  H_{\rm s}
  \mathbf{m}
  \times
  \left(
    \mathbf{p}
    \times
    \mathbf{m}
  \right)
  +
  \alpha 
  \mathbf{m}
  \times
  \frac{d \mathbf{m}}{dt},
  \label{eq:LLG}
\end{equation}
where $\gamma$ and $\alpha$ are 
the gyromagnetic ratio and Gilbert damping constant, respectively. 
The magnetic field is defined as $\mathbf{H}=-\partial E/(\partial M \mathbf{m})$. 
The strength of the spin torque is 
\begin{equation}
  H_{\rm s}
  =
  \frac{\hbar \eta I}{2e MV (1 + \lambda \mathbf{m}\cdot\mathbf{p})},
  \label{eq:H_s}
\end{equation}
where $V$ is the volume of the free layer. 
Two dimensionless parameters, 
$\eta$ and $\lambda$, 
whose ranges are $0<\eta<1$ and $-1<\lambda<1$, 
determine the magnitude of the spin polarization 
and the dependence of the spin torque strength on 
the relative angle between the magnetizations ($\cos^{-1}\mathbf{m}\cdot\mathbf{p}$), respectively. 
The relationships among $\eta$, $\lambda$, and other material parameters 
depend on the theoretical model: 
for example, in the ballistic transport theory in MTJs \cite{slonczewski89,slonczewski02}, 
$\eta$ is proportional to the spin polarization of the density of state of the free layer 
and $\lambda=\eta^{2}$. 
The form of eq. (\ref{eq:H_s}) is common 
for spin torque in not only MTJs but also giant magnetoresistive (GMR) systems \cite{slonczewski96,xiao04}. 
In particular, $\lambda$ plays a key role 
in the magnetization dynamics of this system. 


In the self-oscillation state, 
the energy supply by the spin torque balances with 
the energy dissipation due to the damping, 
and therefore, the magnetization precesses on the constant energy line. 
From eq. (\ref{eq:LLG}), 
the energy change due to the spin torque and the damping is described as 
$d E/dt=-M\mathbf{H}\cdot(d\mathbf{m}/dt)=\mathcal{W}_{\rm s}+\mathcal{W}_{\alpha}$, 
where 
\begin{equation}
  \mathcal{W}_{\rm s}
  =
  \frac{\gamma M H_{\rm s}}{1+\alpha^{2}}
  \left[
    \mathbf{p}
    \cdot
    \mathbf{H}
    -
    (\mathbf{m}\cdot\mathbf{p})
    (\mathbf{m}\cdot\mathbf{H})
    -
    \alpha
    \mathbf{p}
    \cdot
    (\mathbf{m}\times\mathbf{H})
  \right],
  \label{eq:W_s}
\end{equation} 
\begin{equation}
  \mathcal{W}_{\alpha}
  =
  -\frac{\alpha\gamma M}{1+\alpha^{2}} 
  \left[
    \mathbf{H}^{2}
    -
    (\mathbf{m}\cdot\mathbf{H})^{2}
  \right],
  \label{eq:W_alpha}
\end{equation}
are the work done by the spin torque 
and the dissipation due to the damping, respectively \cite{taniguchi13}. 
By assuming that the magnetization tilts from the $z$-axis 
with an angle $\theta=\cos^{-1}m_{z}$ 
and averaging $dE/dt$ over one precession period $\tau$, 
we found that the current $I(\theta)$ satisfying $\overline{dE/dt}=(1/\tau)\oint dt (dE/dt)=0$ is 
\cite{taniguchi13IEEE}
\begin{equation}
\begin{split}
  I(\theta)
  =&
  \frac{2\alpha e \lambda MV}{\hbar \eta \cos\theta}
  \left(
    \frac{1}{\sqrt{1-\lambda^{2}\sin^{2}\theta}}
    -
    1
  \right)^{-1}
\\
  &\times
  \left[
    H_{\rm appl}
    +
    \left(
      H_{\rm K}
      -
      4\pi M
    \right)
    \cos\theta
  \right]
  \sin^{2}\theta.
  \label{eq:I_theta}
\end{split}
\end{equation}
The oscillation frequency at the angle $\theta$ is 
$f(\theta)=1/\tau=\gamma [H_{\rm appl}+(H_{\rm K}-4\pi M)\cos\theta]/(2\pi)$. 
The angle $\theta$ increases with increasing the current, 
which results red-shift of the oscillation frequency \cite{comment1}. 
The critical current for precession is defined as 
$I_{\rm c}=\lim_{\theta \to 0}I(\theta)$, 
and is given by 
\begin{equation}
  I_{\rm c}
  =
  \frac{4\alpha eMV}{\hbar\eta\lambda}
  \left(
    H_{\rm appl}
    +
    H_{\rm K}
    -
    4\pi M
  \right).
  \label{eq:I_c}
\end{equation}
The critical current $I_{\rm c}$ diverges 
in the limit of $\lambda \to 0$ 
because when $H_{\rm s}$ is independent of 
the relative angle of the magnetization, 
and when the equilibrium direction of the magnetization is perpendicular to $\mathbf{p}$, 
the energy supply from the spin torque over one precession period is zero, 
making it impossible for the spin torque to induce the magnetization dynamics. 



\begin{figure}
\centerline{\includegraphics[width=0.5\columnwidth]{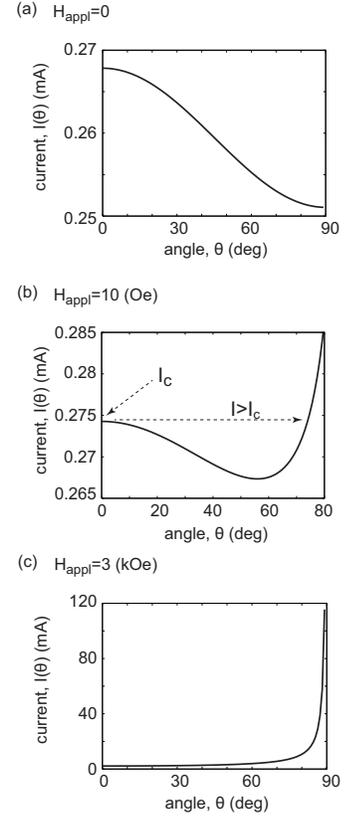}}\vspace{-3.0ex}
\caption{
         Dependences of $I(\theta)$ [eq. (\ref{eq:I_theta})] on 
         the tilted angle of the magnetization $\theta$ 
         for (a) $H_{\rm appl}=0$, (b) $10$, and (c) $3 \times 10^{3}$ Oe, respectively, 
         where $\lim_{\theta\to 0}I(\theta)=I_{\rm c}$. 
         The ranges of $\theta$ in (a) and (c) are $0 \le \theta \le 90^{\circ}$ 
         while that in (b) is $0 \le \theta \le 80^{\circ}$ 
         to emphasize the local minimum of $I(\theta)$. 
         \vspace{-3ex}}
\label{fig:fig2}
\end{figure}



Figures \ref{fig:fig2}(a)-\ref{fig:fig2}(c) show 
the dependences of $I(\theta)$, eq. (\ref{eq:I_theta}), on 
the tilted angle of the magnetization $\theta$ 
with $H_{\rm appl}=0$, $10$, and $3 \times 10^{3}$ Oe, respectively. 
The values of the other parameters are 
$H_{\rm K}=18.6$ kOe, 
$4\pi M=18.2$ kOe, 
$V=\pi \times 60 \times 60 \times 2$ nm${}^{3}$, 
$\eta=0.54$, 
$\lambda=\eta^{2}$, 
$\gamma=17.32$ MHz/Oe, 
and $\alpha=0.005$, 
which are estimated from the experiments \cite{kubota13,konoto13}. 
Depending on the value of the applied field, 
the dependence of $I(\theta)$ on $\theta$ is drastically changed, 
from which the following three distinguishable 
current dependences 
shown in Figs. \ref{fig:fig2}(a)-\ref{fig:fig2}(c) are expected. 

\begin{figure}
\centerline{\includegraphics[width=0.7\columnwidth]{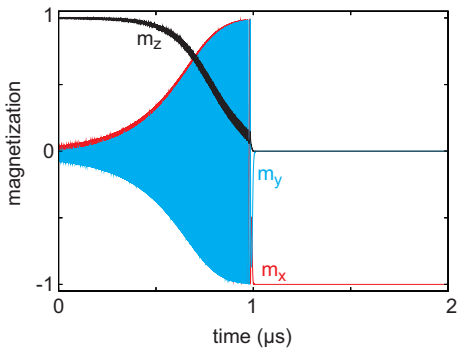}}\vspace{-3.0ex}
\caption{
         Magnetization dynamics in the absence of the applied field, 
         where the red, blue, and black lines correspond to $m_{x}$, $m_{y}$, and $m_{z}$, respectively. 
         The red line ($m_{x}$) below 1 $\mu$s overlaps the blue line. 
         The current magnitude is 0.3 mA ($> I_{\rm c}\simeq 0.27$ mA). 
         \vspace{-3ex}}
\label{fig:fig3}
\end{figure}

First, in the absence of the applied field, 
$I(\theta)$ monotonically decreases as the angle $\theta$ increases, 
and remains finite in the limit of $\theta \to \pi/2$, 
as shown in Fig. \ref{fig:fig2}(a). 
These indicate that, 
once the current magnitude reaches the critical current $I_{\rm c}$, 
the magnetization immediately moves to the film plane ($xy$-plane) 
because $I(0 < \theta < \pi/2)<I_{\rm c}$. 
Figure \ref{fig:fig3} shows the time evolutions of the component of $\mathbf{m}$ 
for $I=0.3 {\rm \ mA} >I_{\rm c} \simeq 0.27 {\rm \ mA}$. 
The magnetization reaches $\theta=\pi/2$, 
and the dynamics stops at $\mathbf{m}=-\mathbf{e}_{x}$ 
because both the field torque and the spin torque, 
which are the first and second terms on the right-hand side of eq. (\ref{eq:LLG}), 
are zero at this point. 
Therefore, self-oscillation cannot be realized 
in the absence of the applied field. 
It should be noted that 
since the spin torque prefers the anti-parallel alignment of the magnetizations for $I>0$, 
$\mathbf{m}$ stops at $-\mathbf{e}_{x}$, 
although all torques are also zero at $+\mathbf{e}_{x}$. 



\begin{figure}
\centerline{\includegraphics[width=0.7\columnwidth]{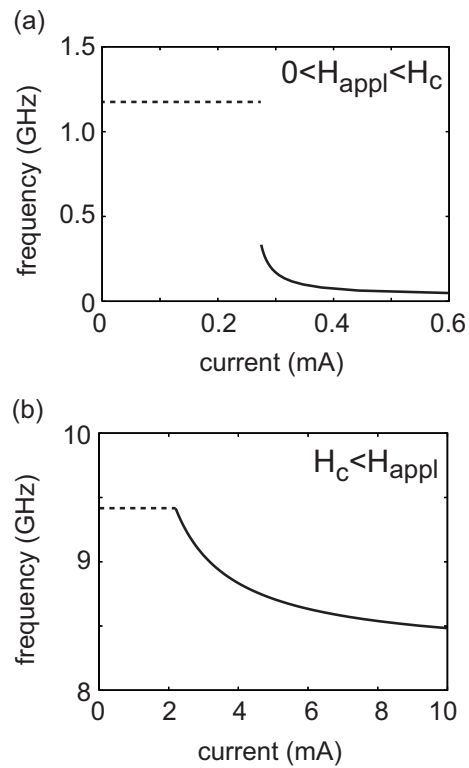}}\vspace{-3.0ex}
\caption{
         The solid lines are the current dependences of the oscillation frequency 
         in the self-oscillation state 
         for (a) $0<H_{\rm appl}<H_{\rm c}$ 
         and (b) $H_{\rm c}<H_{\rm appl}$. 
         The dashed lines are the FMR frequencies. 
         \vspace{-3ex}}
\label{fig:fig4}
\end{figure}

Second, when the magnetic field is smaller than a certain value $H_{\rm c}$, 
i.e., $0<H_{\rm appl}<H_{\rm c}$, 
$I(\theta)$ shows a local minimum, as shown in Fig. \ref{fig:fig2}(b). 
The theoretical formula and the value of $H_{\rm c}$ are derived below. 
In this intermediate region, 
when the current magnitude reaches $I_{\rm c}$, 
the magnetization moves to a certain angle $\theta_{0}$, 
which satisfies $I(\theta_{0})=I_{\rm c}$. 
For example, in Fig. \ref{fig:fig2}(b), 
$\theta_{0} \simeq 74^{\circ}$ for $H_{\rm appl}=10$ Oe. 
By increasing the current magnitude from $I_{\rm c}$, 
the tilted angle $\theta$ continuously increases from $\theta_{0}$. 
The self-oscillation can be realized with the frequency $f(\theta)$. 
It should be noted that, 
below $I_{\rm c}$, the power spectrum of the STO peaks at 
the ferromagnetic resonance (FMR) frequency $f_{\rm FMR}=f(\theta=0)=\gamma (H_{\rm appl}+H_{\rm K}-4\pi M)/(2\pi)$ 
due to the mag-noise effect \cite{kubota13}. 
Since the tilted angle $\theta$ discontinuously changes at $I_{\rm c}$, 
the discontinuity of the oscillation frequency 
as a function of the current is expected, 
as shown in Fig. \ref{fig:fig4} (a). 


Third, $I(\theta)$ monotonically increases as the current increases 
for the applied field satisfying $H_{\rm c}<H_{\rm appl}$. 
In this case, the tilted angle of the magnetization 
continuously increases as the current increases from $I_{\rm c}$. 
Therefore, the oscillation frequency changes from $f_{\rm FMR}$ for the current above $I_{\rm c}$, 
as shown in Fig. \ref{fig:fig4} (b). 



An important assumption used above is that 
the tilted angle of the magnetization, $\theta$, is constant 
during the precession. 
Based on this assumption, 
eq. (\ref{eq:I_theta}) predicts that 
$\lim_{\theta \to \pi/2}I(\theta) = \infty$ for $H_{\rm appl} >0$, 
which means that the magnetization cannot reach the film plane. 
Then, the oscillation frequency saturates to $f(\theta=\pi/2)=\gamma H_{\rm appl}/(2\pi)$ 
in the large current limit. 
Also, the controllable range of the oscillation frequency 
for $H_{\rm c}<H_{\rm appl}$ is $f(\theta=0)-f(\theta=\pi/2)=\gamma (H_{\rm K}-4\pi M)/(2\pi)$, 
which depends on the perpendicular anisotropy only. 
Strictly speaking, however, 
the angle $\theta$ is not a constant during the precession 
because the directions of the spin torque for $m_{x}>0$ and $m_{x}<0$ are opposite, 
as shown in Fig. \ref{fig:fig1}. 
When $\theta$ becomes close to $\pi/2$ by a large current, 
the magnetization can reach the film plane and stops its dynamics 
because the spin torque for $m_{x}<0$ moves 
the magnetization closer to the film plane. 
Therefore, in reality, 
the solid lines in Figs. \ref{fig:fig4}(a) and \ref{fig:fig4}(b) break at a certain current 
above which the self-oscillation cannot be realized. 
The magnitude of such current depends on the applied field magnitude. 
However, the investigation of such current or field magnitude 
requires a breakthrough of the constant $\theta$ assumption, 
and is beyond the scope of this letter. 


The reason why the existence of the applied field determines 
whether the magnetization can reach $\theta=\pi/2$ or not is as follows. 
The tilted angle of the magnetization $\theta$ is determined 
by the balance between the spin torque and the damping. 
In the absence of the applied field, 
the energy dissipation due to the damping, eq. (\ref{eq:W_alpha}), 
rapidly decreases as the angle $\theta$ increases, 
compared with the work done by spin torque, eq. (\ref{eq:W_s}), 
because $\mathcal{W}_{\alpha}$ is on the second order of 
the field $\mathbf{H}=(H_{\rm K}-4\pi M)m_{z} \mathbf{e}_{z}$ 
while $\mathcal{W}_{\rm s}$ is on the first order of $\mathbf{H}$. 
Then, once the spin torque overcomes the damping at $\theta=0$, 
the energy supply from the spin torque is always larger than 
the energy dissipation due to the damping 
during $0 < \theta < \pi/2$. 
Therefore, the magnetization can reach $\theta=\pi/2$. 
Because both the energy supply from the spin torque 
and the energy dissipation due to the damping are zero at $\theta=\pi/2$,  
i.e., $\overline{dE(\theta=\pi/2)/dt}=0$, 
the magnetization dynamics stops at $\theta=\pi/2$. 
However, in the presence of the applied field, 
the damping can balance with the spin torque 
at $0<\theta<\pi/2$ 
because of the presence of the constant term $H_{\rm appl}$ 
in the magnetic field $\mathbf{H}=[H_{\rm appl}+(H_{\rm K}-4\pi M)m_{z}]\mathbf{e}_{z}$. 
Therefore, the self-oscillation of the magnetization 
with the angle $\theta$ can be realized. 
At $\theta=\pi/2$, 
the direction of the spin torque is parallel to the film plane, 
which means that the work done by spin torque is zero. 
On the other hand, 
the energy dissipation due to the damping remains finite, 
i.e., $\overline{dE(\theta=\pi/2)/dt}=-\alpha \gamma M H_{\rm appl}^{2}/(1+\alpha^{2})$. 
Therefore, the magnetization cannot reach $\theta=\pi/2$. 


The value of the critical field $H_{\rm c}$ can be determined by 
the condition in which $dI(\theta)/d\theta > 0$ near $\theta \gtrsim 0$, 
and is given by 
\begin{equation}
  H_{\rm c}
  =
  \frac{3 \lambda^{2}}{2-3\lambda^{2}}
  \left(
    H_{\rm K}
    -
    4\pi M
  \right), 
  \label{eq:critical_field}
\end{equation}
which is 59 Oe for the above parameters. 
We emphasize that $H_{\rm c}$ depends on 
the magnetic anisotropy and the spin torque parameter $\lambda$ only. 
As discussed above, 
when the applied field magnitude is larger than eq. (\ref{eq:critical_field}), 
the controllable range of the oscillation frequency by the current is 
$\gamma(H_{\rm K}-4\pi M)/(2\pi)$. 
On the other hand, when $H_{\rm appl}<H_{\rm c}$, 
the controllable range is suppressed 
because of the discontinuous change of the tilted angle of the magnetization. 


Let us briefly discuss the relation 
between our previous work \cite{kubota13} and this work. 
Reference \cite{kubota13} experimentally investigated 
the current dependence of the oscillation frequency of STO. 
Because the applied field magnitude used in Ref. \cite{kubota13} (typically, 2 kOe) 
is much larger than $H_{\rm c}$, 
the continuous change of the oscillation frequency was observed, 
as shown in Fig. 3 (c) of Ref. \cite{kubota13}. 


At the end of this letter, 
let us mention that 
the situation considered here is similar to 
the switching of the perpendicularly magnetized free layer 
by an in-plane polarized spin current injected by the spin-Hall effect \cite{liu12}. 
Similar to the above discussion, 
in the spin-Hall system, 
the magnetization cannot cross over the film plane 
by the spin torque only, in principle. 
Therefore, to assist or prevent the switching, 
a magnetic field which has a component along the film plane was applied \cite{liu12}. 
In the case of STO discussed in this letter, 
the field-like torque \cite{tulapurkar05,kubota08,sankey08}, 
which is neglected in the above calculation, 
plays a role of a torque due to a magnetic field along the film plane. 
Then, the field-like torque may changes 
the magnetization dynamics, 
especially in the zero-field limit. 
The investigation of the effect of the field-like torque 
will be an important work in future.

In conclusion, 
we studied the oscillation properties of the STO 
consisting of a perpendicularly magnetized free layer 
and an in-plane magnetized pinned layer 
by analyzing the energy balance between the spin torque and the damping. 
We found the existence of the critical value 
of the external magnetic field applied normal to the film plane, $H_{\rm c}$. 
When the applied field is below $H_{\rm c}$, 
the tilted angle of the magnetization discontinuously changes 
above the critical current. 
The controllable range of the oscillation frequency by the current is suppressed 
due to the discontinuity. 
Above $H_{\rm c}$, 
the controllable range of the oscillation frequency is 
$\gamma (H_{\rm K}-4\pi M)/(2 \pi)$. 
The value of the critical field depends on 
the perpendicular magnetic anisotropy, 
the saturation magnetization, 
and the spin torque parameter $\lambda$. 


The authors would like to acknowledge 
H. Naganuma, T. Yorozu, H. Maehara, A. Emura, M. Konoto, A. Fukushima, S. Okamoto, K. Kudo, H. Suto, T. Nagasawa, R. Sato, 
and M. Hayashi 
for their valuable discussions. 




\begin{thebibliography}{10}
\bibitem{comment1}
Blue-shift of the oscillation frequency can be expected when the crystalline
  anisotropy field $H_{\rm K}$ is smaller than the demagnetization field $4\pi M$ and the applied field magnitude is larger than $4\pi M-H_{\rm K}$.
\end{thebibliography}


\end{document}